\documentclass[11pt]{article}
\usepackage{color,graphicx}
\usepackage{amssymb,amsmath}
\usepackage{mathrsfs}
\usepackage{bbm}

\usepackage{authblk}

\allowdisplaybreaks \numberwithin{equation}{section}
\newtheorem{thm}{Theorem}[section]
\newtheorem{prp}[thm]{Proposition}
\newtheorem{lem}[thm]{Lemma}
\newtheorem{dfn}[thm]{Definition}

\newtheorem{example}[thm]{Example}
\newenvironment{exa}{\begin{example} \rm }{ \end{example}}
\newtheorem{remark}[thm]{Remark}

\newenvironment{rmk}{\begin{remark} \rm }{\hfill $\Box$ \end{remark}}
\newenvironment{prf}{\noindent {\it Proof:} \ }{\hfill $\Box$}

\newcommand\od{\mathrm{d}}
\newcommand\p{\partial}\newcommand\pd{\partial}

\newcommand{\nn}{\nonumber}

\newcommand{\ld}{\lambda}

\newcommand{\dt}{\delta}
 
\newcommand{\ka}{\kappa}
\newcommand{\vp}{\varphi}

\newcommand{\res}{\mathrm{res}\,}

\newcommand\Z{\mathbb{Z}}

\newcommand\cA{\mathcal{A}}
\newcommand\cD{\mathcal{D}}

\newcommand\cM{\mathcal{M}}

\newcommand{\set}[1]{\left\{#1\right\}}

\newcommand{\bm}[1]{\mathbf{#1}}
\newcommand{\bt}{\bm{t}}  \newcommand{\hbt}{\hat{\bm{t}}}

\begin{document}

\title{Bilinear Equation and Additional Symmetries for an Extension of the Kadomtsev--Petviashvili Hierarchy}
\author[]{Jiaping Lu}
\author[]{ Chao-Zhong Wu%\footnote{Corresponding Author}
}
\affil[]{School of Mathematics,  Sun
Yat-Sen University,
Guangzhou 510275, P.R. China 
%\\
%Email: jiapingljp@hotmail.com; wuchaozhong@sysu.edu.cn
}

\date{}
\maketitle

\begin{abstract}
An extension of the Kadomtsev--Petviashvili (KP) hierarchy defined via scalar pseudo-differential operators was studied in \cite{SB,WZ}.
In this paper, we represent the extended KP hierarchy into the form of bilinear equation of (adjoint) Baker--Akhiezer functions, and construct its additional symmetries. As a byproduct, we derive the Virasoro symmetries for the constrained KP hierarchies.
\end{abstract}

\parskip 1ex

\section{Introduction }

As a fundamental model in the theory of integrable systems, the Kadomtsev--Petviashvili (KP) hierarchy can be defined as follows. Let
\begin{equation} \label{Lckp}
L_{KP}=\p+v_1\p^{-1}+v_2\p^{-2}+\dots, \quad \p=\frac{\od}{\od x},
\end{equation}
be a pseudo-differential operator whose coefficients $v_i$ are scalar unknown functions of the spacial coordinate $x$, then the KP hierarchy is composed by the following evolutionary equations of $v_i$ as
\begin{equation}\label{kp}
 \frac{\p {L_{KP}}}{\p t_k}=\left[({L_{KP}}^k)_+, {L_{KP}}\right],\quad
 k=1,2,3,\dots.
\end{equation}
Here and below the subscript ``$+$'' of a pseudo-differential operator means to take its purely differential part, while the subscript ``$-$'' means to take its negative part. Suppose that the equations \eqref{kp} are imposed with the constraint $({L_{KP}}^n)_-=0$ with some integer $n\ge2$, then they form the Gelfand--Dickey (or the $(n-1)$-th Korteweg--de Vries) hierarchy.

It is known that the KP hierarchy \eqref{kp} possesses a series of
bi-Hamiltonian structures derived by the $R$-matrix formalism
\cite{STS}, and that it can be represented into the form of bilinear
equation of (adjoint) Baker--Akhiezer functions or of a tau function
\cite{DKJM-KPBKP}. Such bi-Hamiltonian structures and the bilinear
equation can be reduced to that of the Gelfand--Dickey hierarchies.
What is more, for the KP hierarchy there is a class of
non-isospectral symmetries named as the additional symmetries that
can be constructed via the so-called Orlov--Schulman operators
\cite{OS}. The flows of such additional symmetries commute with all the
time flows $\p/\p t_k$ but do not commute between themselves;
instead, they generate the $W_{1+\infty}$ algebra. As an
application of the additional symmetries for the KP
hierarchy, part of these symmetries can be reduced to the Virasoro symmetries
for its subhierarchies such like the Gelfand--Dickey hierarchies. Note that Virasoro symmetries reveal crucial
properties of a large amount of integrable hierarchies including the
Gelfand--Dickey hierarchies and the Drinfeld--Sokolov hierarchies,
see e.g. \cite{AvM, Di-add1, Di-add2, Wu-DS} and references therein.

In the definition of the KP hierarchy \eqref{kp}, it is used pseudo-differential operators with only finitely many positive powers in $\p$.
The notion of pseudo-differential operator was generalized in \cite{LWZ} to be over a certain graded differential algebra $\cA$ such that these operators may contain infinitely many positive powers in $\p$ (see Section~2 below for details). By using these operators, in \cite{WZ} Zhou and one of the authors of the present paper considered an integrable hierarchy, which can be viewed as a subhierarchy of the dispersionful analogue \cite{SB} of the universal Whitham hierarchy. More exactly, let
\begin{equation} \label{PhP}
P= \p+\sum_{i\ge1}u_i  (\p-\vp)^{-i}, \quad \hat{P}=
(\p-\vp)^{-1}\hat{u}_{-1}+\sum_{i\ge0}\hat{u}_i (\p-\vp)^i
\end{equation}
with $u_i, \hat{u}_i$ and $\vp$ being certain unknown functions belong to $\cA$, then the
following evolutionary equations are well defined:
\begin{align}\label{ekp1}
& \frac{\p }{\p t_k}(P,\hat{P})=\left([(P^k)_+, P],[(P^k)_+,
\hat{P}]\right), \quad \frac{\p}{\p \hat{t}_k}(P,\hat{P})=\left([-(\hat{P}^k)_-, P],
[-(\hat{P}^k)_-, \hat{P}]\right),
\end{align}
where $k=1, 2, 3, \dots$. These evolutionary equations compose an integrable hierarchy, which is named as the extended KP hierarchy. In fact, this hierarchy is reduced to the KP hierarchy \eqref{kp} whenever $\hat{P}=0$ (note that the operator $P$ gives an alternative expression of $L_{KP}$). On the other hand, if one lets $\vp\to0$ and impose certain B-type symmetry conditions to the operators $P$ and $\hat{P}$, then the flows in \eqref{ekp1} with $k\in\Z_{>0}^{\mathrm{odd}}$ give the two-component BKP (2-BKP) hierarchy \cite{DJKM-KPtype, LWZ}. With the $R$-matrix method applied in the cases of the KP and the 2-BKP hierarchies \cite{STS, Wu-Rmat}, the extended KP hierarchy \eqref{ekp1} was shown to possess infinitely many bi-Hamiltonian structures \cite{WZ}.

In this paper we assume $\vp=\p(f)$ in \eqref{PhP} with some homogeneous function $f\in\cA$ of degree $0$. We will show that the operators $P$ and $\hat P$ can be represented in a dressing form as
\begin{equation} \label{PPh0}
P=\Phi \p\Phi^{-1},\quad \hat{P}=\hat{\Phi}
\p^{-1}\hat{\Phi}^{-1},
\end{equation}
where $\Phi$ and $\hat\Phi$ are pseudo-differential operators of the form
\begin{align*}
&\Phi=1+\sum_{i\ge 1}a_i \p^{-i},\quad
\hat{\Phi}=e^f \left(1+\sum_{i\ge 1}b_i \p^{i}\right).
\end{align*}
With the help of these two dressing operators in the extended KP hierarchy, we will introduce two Baker--Akhiezer functions $\psi(\bt, \hbt; z)$, $\hat\psi(\bt, \hbt; z)$ and their adjoints $\psi^\dag(\bt, \hbt; z)$, ${\hat\psi}^\dag(\bt, \hbt; z)$ that depend on the time variables $\bm{t}=(t_1,t_2,t_3,\dots)$,
$\hat{\bm{t}}=(\hat{t}_1,\hat{t}_2,\hat{t}_3,\dots)$, and a nonzero parameter $z$. Our first main result is the following theorem (see Theorem~\ref{thm-bl} for a more precise version).
\begin{thm}\label{thm-bl0}
The extended KP hierarchy \eqref{ekp1} can be represented equivalently to a bilinear equation as
\begin{equation}\label{bl}
\res_{\!z}\left(\psi(\bt, \hbt; z)\psi^\dag(\bt', \hbt'; z)\right)=\res_{\!z}\left(\hat{\psi}(\bt, \hbt; z)\hat{\psi}^\dag(\bt', \hbt'; z)\right)
\end{equation}
with arbitrary time variables $(\bt, \hbt)$ and $(\bt', \hbt')$.
\end{thm}

Our second main result is the construction of  additional symmetries
for the extended KP hierarchy, with the help of  certain
Orlov--Schulman operators given by the above dressing operators
$\Phi$ and $\hat\Phi$. Such additional symmetries will be
shown to generate a $W_{1+\infty}\times W_{1+\infty}$ algebra. These
results will be applied to study the $(n,1)$-constrained KP
hierarchy (see, e.g., \cite{ANP, BX, Cheng-ncKP,
Di, Kr}), which is denoted as $\hbox{cKP}_{n,1}$ and can be reduced from the extended KP hierarchy \eqref{ekp1}
under the constraint
\[
P^n=\hat{P}
\]
with a given integer $n\ge1$. In fact, the Virasoro symmetries for the  $\hbox{cKP}_{n,1}$ hierarchy (even for more general cases) were proposed by Aratyn, Nissimov and Pacheva \cite{ANP}, with the method of adding certain ``ghost'' symmetry flows such that some nonlocal actions on functions are involved. They also showed that, the  $\hbox{cKP}_{n,1}$ hierarchy subject to the subsidiary condition of invariance under the lowest Virasoro symmetry flow can be applied to compute explicit Wronskian solution for the two-matrix model partition function  \cite{ANP-DB}. As to be seen, the Virasoro symmetries appearing in \cite{ANP, ANP-DB} for the  $\hbox{cKP}_{n,1}$ hierarchy can be obtained in an alternative way, starting from the additional symmetries for the extended KP hierarchy (see Proposition~\ref{thm-ANP} below).

This paper is arranged as follows. In the next section, we will recall the pseudo-differential operators of the first and the second types over a graded differential algebra. In Section~3, we will recall the definition of the extended KP hierarchy, and represent it into the form of a bilinear equation. In Section~4, we will construct the additional symmetries for the extended KP hierarchy, and then study the Virasoro symmetries for the $\hbox{cKP}_{n,1}$ hierarchy. The final section is devoted to some remarks.

\section{Pseudo-differential operators}\label{sec-eKP}

%\subsection{Preliminary notations}
Let $\cA$ be a commutative associative algebra, and $\p: \cA\to\cA$
be a derivation. We consider the linear space $\cD(\cA)=\set{\sum_{i\in\Z}f_i\p^i\mid f_i\in\cA}$ and its subsets. For instance, the set of pseudo-differential operators is
\[
\cA((\p^{-1}))=\set{\sum_{i\le k}f_i\p^i\mid f_i\in\cA, k\in\Z},
\]
and it becomes an associative algebra if a product is defined by
\begin{equation}\label{pro}
f \p^i\cdot g \p^j=\sum_{r\geq0}\binom{i}{r}f\, \p^r(g)\,
\p^{i+j-r}, \quad f,\,g\in\cA.
\end{equation}
For any two  pseudo-differential operators $A$ and $B$, their
commutator means $[A,B]=A B-B A$. Clearly, one has $[\p, f]=\p(f)$
for any $f\in\cA$.

In the present paper we assume the algebra $\cA$ to be a graded one.
Namely, $\mathcal{A}=\prod_{i\geq0}\cA_i$, such that
\[
\cA_i\cdot\cA_j\subset\cA_{i+j}, \quad \p(\cA_i)\subset\cA_{i+1}.
\]
Denote $\cD^-=\cA((\p^{-1}))$, which is called the algebra of
pseudo-differential operators of the first type over $\cA$. In
contrast, by the algebra of pseudo-differential operators of the
second type over $\cA$ it means \cite{LWZ}
\begin{equation}\label{Dpl}
\cD^+=\left\{ \sum_{i\in\Z}\sum_{j\ge \max\{0,k-i\}}a_{i,j} \p^i
\mid a_{i,j}\in\cA_j, k\in\Z \right\},
\end{equation}
whose product is also defined by \eqref{pro}. One observes that an
operator in $\cD^+$ may contain infinitely many positive powers in
$\p$.

Given an element $A=\sum_{i\in\Z} {f}_i \p^i\in\cD(\cA)$, its differential part, negative part and residue are defined
respectively as:
\begin{equation}\label{}
A_+=\sum_{i\ge0} f_i \p^i, \quad A_-=\sum_{i<0} f_i \p^i, \quad \res
A=f_{-1}.
\end{equation}
It is easy to see that
\[
\left(\cD^\mp\right)_\pm\subset \cD^-\cap\cD^+.
\]
What is more, on each $\cD^\mp$ there is an anti-automorphism
defined by
\[
\p^*=-\p, \quad f^*=f \hbox{ with } f\in\cA.
\]
Clearly, for any $A\in\cD^\mp$, one has
\[
\res A^*=-\res A.
\]

In what follows we will use the notation $\cA_{\ge r}=\prod_{i\ge
r}\cA_i$ with $r\in\Z_{\ge0}$. Given an element $\vp\in\cA_{\ge1}$,
the following two maps are well defined with $\p$ replaced by
$\p-\vp$, that is,
\begin{align}
\mathscr{S}_\vp: \quad \cD^\mp&\to\cD^\mp, \nn\\
 \sum {f}_i \p^i &\mapsto \sum {f}_i (\p-\vp)^i.
\label{Svp}
\end{align}
For instance, we have
\begin{align}\label{}
\mathscr{S}_\vp(\p^{-1})=&(\p-\vp)^{-1}=\p^{-1}(1-\vp\p^{-1})^{-1} \nn\\
=&\p^{-1}+\p^{-1}\vp\p^{-1} +\p^{-1}\vp\p^{-1}\vp\p^{-1}+\dots.
\end{align}
In \cite{WZ}, it was verified that the maps $\mathscr{S}_\vp$ are
automorphisms on each $\cD^\mp$. Accordingly, the algebras $\cD^\mp$
can be represented as follows:
\begin{align}\label{Dvpm}
\cD^-=&\set{\sum_{i\le k}g_i(\p-\vp)^i\mid g_i\in\cA, k\in\Z},\\
 \cD^+=&\left\{
\sum_{i\in\Z}\sum_{j\ge \max\{0,k-i\}}b_{i,j} (\p-\vp)^i \mid
b_{i,j}\in\cA_j, k\in\Z \right\}, \label{Dvpp}
\end{align}
and their product can be defined equivalently by
\[
f (\p-\vp)^i\cdot g (\p-\vp)^j=\sum_{r\geq0}\binom{i}{r}f\,
\p^r(g)\, (\p-\vp)^{i+j-r}, \quad f,\,g\in\cA.
\]
Moreover, it is easy to verify the following properties: for any
$A\in\cD^\mp$,
\begin{equation}\label{Spres}
(\mathscr{S}_\vp A)_\pm=\mathscr{S}_\vp(A_\pm), \quad \res (\mathscr{S}_\vp A)=\res A, \quad (\mathscr{S}_\vp A)^*=\mathscr{S}_{-\vp}A^*.
\end{equation}

\section{The extended KP hierarchy and its bilinear equation}

We proceed to recall the definition of the extended KP hierarchy,
and then  capsule it into a bilinear equation of certain (adjoint)
Baker--Akhiezer functions.

\subsection{The extended KP hierarchy}
Let ${\cM}$ be an infinite-dimensional manifold with
coordinate
\[
\bm{a}=(a_1, a_2, a_3, \dots; f, b_1, b_2, b_3, \dots ).
\]
We consider the following graded
algebra of formal differential polynomials:
\begin{equation}\label{tA}
{\cA}=C^\infty(S^1\to{\cM})[[\p^r(\bm{a})\mid r\ge1]],
\end{equation}
in which the derivation is $\p=\od/\od x$ with $x$ being the
coordinate of the loop $S^1$, and
\[
\deg \bm{a}=0, \quad \deg \p^r(\bm{a})=r.
\]
Over the graded differential algebra $\cA$, it is defined the algebras
$\cD^-$ and  $\cD^+$ of pseudo-differential operators of the first
type and of the second type, respectively.

Let us consider two pseudo-differential operators as follows:
\begin{align} \label{Phi}
&\Phi=1+\sum_{i\ge 1}a_i \p^{-i}\in\cD^-,\\
&\hat{\Phi}=e^f \left(1+\sum_{i\ge 1}b_i \p^{i}\right)\in\cD^+. \label{Phih}
\end{align}
These two operators have inverses of the form
\begin{align} \label{Phiinv}
&\Phi^{-1}=1+\sum_{i\ge 1}\tilde{a}_i \p^{-i}\in\cD^-,\\
&\hat{\Phi}^{-1}=\left(1 +\sum_{i\ge 1}\tilde{b}_i
\p^{i}\right)e^{-f} \in\cD^+. \label{Phihinv}
\end{align}
In fact, by expanding $\Phi^{-1}\Phi=1$ and $\hat{\Phi}^{-1}\hat{\Phi}=1$, one sees that the coefficients $\tilde{a}_i$ and $\tilde{b}_i$ are represented as
\[
\tilde{a}_i=-a_i+g_i(a_1,a_2,\dots,a_{i-1}), \quad \tilde{b}_i=-b_i+h_i(b_1,b_2, b_3, \dots)
\]
with $g_i, h_i\in\cA_{\ge1}$.
For instance, one has $\tilde{a}_1=-a_1$, and that $\tilde{b}_1$ is given recursively by
\[
\left.\tilde{b}_1\right|_{\cA_0}=-b_1, \quad
\left.\tilde{b}_1\right|_{\cA_j}=- \sum_{r=1}^{j}b_r\left(
\left.\tilde{b}_1\right|_{\cA_{j-r}}\right)^{(r)} ~~ \hbox{ for } ~~
j\ge1.
\]
Here $\left.\tilde{b}_1\right|_{\cA_j}$ means the degree-$j$ component of $\tilde{b}_1$.

Now let us introduce two pseudo-differential operators as follows:
\begin{equation} \label{PPh}
P=\Phi \p\Phi^{-1}\in\cD^-,\quad \hat{P}=\hat{\Phi}
\p^{-1}\hat{\Phi}^{-1}\in\cD^+.
\end{equation}
\begin{prp}
The operators $P$ and $\hat{P}$ given above can be represented in
the form:
\begin{align}\label{PPh2}
P=\p+\sum_{i\ge1}v_i\p^{-i}, \quad
\hat{P}=(\p-f')^{-1}\rho+\sum_{i\ge0}\hat{v}_i\p^i,
\end{align}
where $v_{i+1}, \hat{v}_i\in \cA_{\ge1}$ for $i\in\Z_{\ge0}$, and
\begin{equation}\label{rho}
\rho=e^{f}\left(\hat\Phi^{-1}\right)^*(1).
\end{equation}
\end{prp}
\begin{prf}
The representation of $P$ is well known, so let us verify the case
of $\hat{P}$. To this end, firstly let us check that $\hat{P}$ given
in \eqref{PPh} takes the form
\begin{equation}\label{}
\hat{P}=(\p-f')^{-1}\left( \sum_{i\ge0}\tilde{v}_i(\p-f')^i\right), \quad
\tilde{v}_i-\dt_{i 0}\in \cA_{\ge1}.
\end{equation}
We substitute this expansion and \eqref{Phih} into an equivalent
version of $\hat{\Phi}\p^{-1}\hat{\Phi}^{-1}=\hat{P}$, say,
\[
e^{-f}(\p-f')\hat{\Phi}\p^{-1}=e^{-f}(\p-f')\hat{P}\hat{\Phi}.
\]
Note $e^{-f}(\p-f')e^{f}=\p$, then we have
\begin{align} \label{eqvtil}
1+b_1'+\sum_{i\ge1}(b_i+b_{i+1}')\p^{i}=\left(
\sum_{i\ge0}\tilde{v}_i\p^i\right)\left(1+ \sum_{i\ge1}b_i\p^i\right),
\end{align}
in which the coefficients of $\p^i$ lead to
\begin{align}\label{}
i=0 &: \quad 1+b_1'=\tilde{v}_0, \label{vtil-1} \\
i\ge1 &: \quad b_i+b_{i+1}'=\tilde{v}_i+\sum_{r=1}^i\sum_{s\ge 0}
\binom{i-r+s}{s}\tilde{v}_{i-r+s}b_{r}^{(s)}. \label{vtil}
\end{align}
The equations \eqref{vtil} yield
\[
b_i=\left.\tilde{v}_i\right|_{\cA_0}+\sum_{r=1}^i \left.\tilde{v}_{i-r}\right|_{\cA_0} b_r, \quad i\ge1,
\]
which together with \eqref{vtil-1} gives
\[
\left.\tilde{v}_i\right|_{\cA_0}=\dt_{i 0}, \quad i\ge0.
\]
The equations \eqref{vtil} also yield
\[
b_{i+1}'=\left.\tilde{v}_i\right|_{\cA_1}+\sum_{r=1}^i \left.\tilde{v}_{i-r}\right|_{\cA_1} b_r, \quad i\ge1,
\]
which together with \eqref{vtil-1} determine $\left.\tilde{v}_i\right|_{\cA_1}$ recursively. In fact, we can obtain a generating function for
$\left.\tilde{v}_i\right|_{\cA_1}$ of a parameter $z$ as
\[
\sum_{i\ge0} \left.\tilde{v}_i\right|_{\cA_1} z^{i+1}=\p \log\left(1+\sum_{i\ge 1}b_i z^i\right).
\]
For $j\ge2$, it follows from \eqref{vtil} that
\[
\left.\tilde{v}_i\right|_{\cA_j}+\sum_{r=1}^i\sum_{s=0}^{j-1} \binom{i-r+s}{s}\left.\tilde{v}_{i-r+s}\right|_{\cA_{j-s}}\cdot b_{r}^{(s)}=0, \quad i\ge1,
\]
hence $\left.\tilde{v}_i\right|_{\cA_j}$ are determined recursively.
By using \eqref{Svp}, we derive that the operator $\hat P$ takes the form \eqref{PPh2}, in which
\begin{equation}\label{}
\rho=\res \left(\hat{\Phi}\p^{-1}\hat{\Phi}^{-1}\right)=\res \left(e^f \p^{-1}\hat{\Phi}^{-1}\right)=e^f \left(\hat\Phi^{-1}\right)^*(1).
\end{equation}
Therefore the proposition is proved.
\end{prf}

\begin{rmk}
From \eqref{rho} and \eqref{Phihinv} it follows that
\begin{equation}\label{}
\rho=e^f \left(\hat\Phi^{-1}\right)^*(1)=1+\sum_{i\ge1}(-1)^i
\tilde{b}_i^{(i)},
\end{equation}
which implies $\rho-1\in\cA_{\ge1}$.
One also sees that, the coefficients of $P$ and $\hat{P}$ have different degrees from that in \cite{WZ}, since now the graded differential algebra $\cA$ is chosen differently.
\end{rmk}

With the help of the operators $P$ and $\hat{P}$, let us define a class of evolutionary equations on $\cM$: for $k\in\Z_{>0}$,
\begin{align}
&\frac{\pd \Phi}{\pd t_k}=- (P^k)_-\Phi, \quad
\frac{\pd \hat{\Phi}}{\pd t_k}=\bigl((P^k)_+ -\dt_{k1} \hat{P}^{-1}\bigr)\hat{\Phi}, \label{ppt1}\\
&\frac{\pd \Phi}{\pd \hat{t}_k}=- (\hat{P}^k)_-\Phi, \quad \frac{\pd
\hat{\Phi}}{\pd \hat{t}_k}=(\hat{P}^k)_+\hat{\Phi}, \label{ppt2}
\end{align}
Here we note that the right hand sides make sense since the operators $(P^k)_+, (\hat{P}^k)_-\in\cD^-\cap\cD^+$, and we assume that the flows $\p/\p t_k$ and $\p/\p\hat{t}_k$ commutate with $\p/\p x$. In particular, it can be seen $\p/\p t_1=\p/\p x$, so in what follows we will just take $t_1=x$.

\begin{prp}\label{thm-PPht}
The flows \eqref{ppt1}, \eqref{ppt2} satisfy, for $k\in\Z_{>0}$,
\begin{align}\label{PPht}
& \frac{\pd P}{\pd t_k}=\left[(P^k)_+, P\right], \quad \frac{\pd \hat{P}}{\pd
t_k}=\left[(P^k)_+, \hat{P}\right], \\
\label{PPhth} & \frac{\pd P}{\pd \hat{t}_k}=\left[-(\hat{P}^k)_-, P\right],
\quad \frac{\pd \hat{P}}{\pd \hat{t}_k}=\left[-(\hat{P}^k)_-, \hat{P}\right].
\end{align}
Moreover, these flows commute with each other.
\end{prp}

\begin{prf}
The proposition can be verified case by case. For instance, we have
\begin{align*}
\frac{\p\hat{P}}{\p t_k}= &\left[\frac{\p\hat\Phi}{\p t_k}{\hat\Phi}^{-1},{\hat P} \right]
= \left[(P^k)_+ -\dt_{k1} \hat{P}^{-1},{\hat P} \right]=\left[(P^k)_+ ,{\hat P} \right], \\
\left[\frac{\p}{\p t_k},\frac{\p}{\p{\hat t}_l} \right]{\hat\Phi}=& \frac{\p}{\p t_k}\left( (\hat{P}^l)_+\hat{\Phi}\right)-\frac{\p}{\p {\hat t}_l}\left( \bigl((P^k)_+ -\dt_{k1} \hat{P}^{-1}\bigr)\hat{\Phi} \right) \\
=& \left[(\hat{P}^l)_+, (P^k)_+ -\dt_{k1} \hat{P}^{-1} \right]\hat{\Phi}+ \left[(P^k)_+, \hat{P}^l\right]_+\hat{\Phi} \\
&- \left(\left[- (\hat{P}^l)_-, P^k\right]_+-\dt_{k1}\left[- (\hat{P}^l)_-, \hat{P}^{-1} \right] \right)\hat{\Phi} \\
=& \left[(\hat{P}^l)_+, (P^k)_+  \right]\hat{\Phi}-\left[ \hat{P}^l , (P^k)_+  \right]_+\hat{\Phi} \\
&+\left[(\hat{P}^l)_-, (P^k)_+  \right]_+\hat{\Phi} -\dt_{k1}\left[ \hat{P}^l , \hat{P}^{-1}  \right]\hat{\Phi} \\
=&0.
\end{align*}
The other cases are almost the same. So we complete the proof.
\end{prf}

The system of equations \eqref{PPht}, \eqref{PPhth} was studied in
\cite{WZ} by Zhou and one of the authors (cf. \cite{SB}), with the
set of unknown functions as
\[
\left\{ v_{i+1}, \hat{v}_i \mid i\in\Z_{\ge0}\right\}\cup\{\rho, \vp=f'\}.
\]
This system is called the extended KP hierarchy for the reason that the flows $\p P/\p t_k$ in \eqref{PPht} with $k\in\Z_{>0}$ compose the well-known KP hierarchy. By virtue of the above proposition, we will also call the system of equations \eqref{ppt1}, \eqref{ppt2} the extended KP hierarchy.
\begin{exa}
From the extended KP hierarchy one can write down some equations explicitly as follows:
\begin{align}
\frac{\p^2 v_1}{\p {t_2}^2}=&\left(\frac{4}{3}\frac{\p v_1}{\p t_3}-4v_1 v_1' -\frac{1}{3}v_1''' \right)', \quad
\frac{\p f}{\p t_2}=2 v_1+(f')^2+f'', \quad
\frac{\p \rho}{\p t_2}= \left(2\rho f'-\rho'\right)'. \label{ekpt2}
\end{align}
\end{exa}

\begin{exa}
By using \eqref{Phih} and \eqref{ppt1} we have, for $k\in\Z_{>0}$,
\begin{align}
\frac{\p f}{\p t_k}=& e^{-f}\frac{\p\hat\Phi}{\p t_k}(1)
=e^{-f}\bigl((P^k)_+ -\dt_{k1} \hat{P}^{-1}\bigr)\hat{\Phi}(1)
= e^{-f}(P^k)_+  e^f (1)  \nn \\
=&\res \left(e^{-f}P^k   e^f\p^{-1}\right) =\res \mathscr{S}_{-f'}\left(P^k  e^f\p^{-1}e^{-f}\right)
\nn \\
=&\res \left(P^k  e^f\p^{-1}e^{-f}\right)= \res \left( P^k (\p-f')^{-1}\right). \label{ft}
\end{align}
Here in the fifth equality we have used \eqref{Spres}. With the same method, we obtain
\begin{align}
\frac{\p f}{\p\hat{t}_k}= \res \left( \hat{P}^k (\p-f')^{-1}\right), \quad k\in\Z_{>0}. \label{ft}
\end{align}
Moreover, for $\dot{t}_k=t_k$ or $\hat{t}_k$ (correspondingly, $\dot{P}=P$ or $\hat{P}$), we have
\begin{align}
\frac{\p\rho}{\p\dot{t}_k}=&\sum_{i\ge1}(-1)^{i-1}\p^i\left(\rho\,\res\,
\dot{P}^k(\p-f')^{-i-1}\right). \label{rhot}
\end{align}
By letting $\vp=f'$, we recover the equations (2.23) and (2.24) in \cite{WZ}.
\end{exa}

\subsection{Baker--Akhiezer functions and the bilinear equation}

%For the extended KP hierarchy, we proceed to introduce its Baker--Akhiezer functions and the bilinear equation they satisfy.

With $z$ being a nonzero parameter, we assign $\p^i\left( e^{x
z}\right)=z^i e^{x z}$ for any $i\in\Z$, and more generally,
\[
\p^{i}\left(g e^{x z}\right)=(\p^{i}g)\left( e^{x z}\right), \quad g\in\cA, ~~ i\in\Z.
\]
Namely, it is the usual action of a differential operator on a
function whenever $i\ge0$, and the integral constants are fixed in a
special way whenever $i<0$. In order to simplify notations, for any
$A\in\cD^\pm$ and exponential functions of the the form $e^{\pm x
z}$, we will just write $A e^{\pm x z}$ instead of $A\left( e^{\pm x
z}\right)$.

Denote $\bm{t}=(t_1=x,t_2,t_3,\dots)$  and
$\hat{\bm{t}}=(\hat{t}_1,\hat{t}_2,\hat{t}_3,\dots)$, and let $\xi$ be given by
\[
\xi(\bm{t}; z)=\sum_{k\in\Z_{>0}}
t_k z^k.
\]
Given a solution of the extended KP hierarchy \eqref{ppt1}, \eqref{ppt2}, let us introduce two
Baker--Akhiezer functions:
\begin{align}\label{wavef}
\psi(\bm{t}, \hat{\bm{t}}; z)=\Phi e^{\xi(\bm{t};z)}, \quad
\hat{\psi}(\bm{t}, \hat{\bm{t}}; z)=\hat{\Phi}  e^{x
z-\xi(\hat{\bm{t}};z^{-1})},
\end{align}
and two adjoint Baker--Akhiezer functions:
\begin{align}\label{baker}
\psi^\dag(\bm{t}, \hat{\bm{t}}; z)=\left(\Phi^{-1}\right)^*  e^{-\xi(\bm{t};z)} , \quad
\hat{\psi}^\dag(\bm{t}, \hat{\bm{t}}; z)=\left(\hat{\Phi}^{-1}\right)^*  e^{-x
z+\xi(\hat{\bm{t}};z^{-1})}.
\end{align}
When there is no confusion, we will just write $\eta(z)=\eta(\bm{t}, \hat{\bm{t}}; z)$   with $\eta\in\{\psi, \hat{\psi}, \psi^\dag, \hat{\psi}^\dag\}$. Based on \eqref{ppt1} and \eqref{ppt2},
it is straightforward to verify the the following Lemma.
\begin{lem}\label{thm-bakert}
The (adjoint) Baker--Akhiezer functions given above satisfy:
\begin{align}\label{}
\mathrm{(i)} \qquad & P\psi(z)=z\psi(z), \quad P^*\psi^\dag(z)=z\psi^\dag(z),  \\
& \hat{P}{\hat \psi}(z)=z^{-1}{\hat \psi}(z), \quad {\hat P}^*{\hat \psi}^\dag(z)=z^{-1}{\hat\psi}^\dag(z); \\
\mathrm{(ii)} \qquad
& \frac{\p\dot\psi(z)}{\p t_k}=\left(P^k\right)_+{\dot\psi(z)}, \quad \frac{\p\dot\psi(z)}{\p \hat{t}_k}=-\left(\hat{P}^k\right)_-\dot\psi(z),  \\
& \frac{\p\dot\psi^\dag(z)}{\p t_k}=-\left(P^k\right)_+^*{\dot\psi^\dag(z)}, \quad \frac{\p\dot\psi^\dag(z)}{\p \hat{t}_k}=\left(\hat{P}^k\right)_-^*\dot\psi^\dag(z)
\end{align}
where $\dot\psi\in\left\{\psi, \, \hat{\psi}\right\}$ and  $\dot\psi^\dag\in\left\{\psi^\dag, \, \hat{\psi}^\dag\right\}$.
\end{lem}

For any formal series $\sum_{i\in\Z}g_i z^i$, its residue is defined by
\[
\res_{\!z} \sum_{i\in\Z}g_i z^i=g_{-1}.
\]
The following lemma is useful.
\begin{lem}[see, for example, \cite{DKJM-KPBKP}] \label{thm-res}
For any pseudo-differential operators $Q, R\in\cD^\pm$, the following equality holds true whenever both sides make sense:
\begin{equation}\label{res}
\res_{\!z}\left(Q e^{z x}\cdot R^* e^{-z x}\right)=\res (Q R).
\end{equation}
\end{lem}

\begin{thm}\label{thm-bl}
The (adjoint) Baker--Akhiezer functions of the extended KP hierarchy satisfy
the following bilinear equation
\begin{equation}\label{bl}
\res_{\!z}\left(\psi(\bt, \hbt; z)\psi^\dag(\bt', \hbt'; z)\right)=\res_{\!z}\left(\hat{\psi}(\bt, \hbt; z)\hat{\psi}^\dag(\bt', \hbt'; z)\right),
\end{equation}
for arbitrary time variables $(\bt, \hbt)$ and $(\bt', \hbt')$. Conversely, suppose that four  functions of the form
\begin{align}\label{wavef2}
&\psi(\bm{t}, \hat{\bm{t}}; z)=\left(1+\sum_{i\ge1}a_i(\bm{t}, \hat{\bm{t}})z^{-i}\right) e^{\xi(\bm{t};z)},  \\
&\hat{\psi}(\bm{t}, \hat{\bm{t}}; z)=e^{f(\bm{t}, \hat{\bm{t}}) }\left(1+\sum_{i\ge1}b_i(\bm{t}, \hat{\bm{t}})z^{i}\right) e^{x
z-\xi(\hat{\bm{t}};z^{-1})},  \\
&\psi^\dag(\bm{t}, \hat{\bm{t}}; z)=\left(1+\sum_{i\ge1} {a}_i^\dag(\bm{t}, \hat{\bm{t}})z^{-i}\right)   e^{-\xi(\bm{t};z)}, \\
&\hat{\psi}^\dag(\bm{t}, \hat{\bm{t}}; z)=e^{-f(\bm{t}, \hat{\bm{t}}) }\left(1+\sum_{i\ge1}{b}_i^\dag(\bm{t}, \hat{\bm{t}})z^{i}\right)  e^{-x
z+\xi(\hat{\bm{t}};z^{-1})} \label{wavef2a}
\end{align}
satisfy the bilinear equation \eqref{bl}, then they are the  Baker--Akhiezer functions and the adjoint  Baker--Akhiezer functions of the extended KP hierarchy.
\end{thm}
\begin{prf}
As a preparation, we introduce the set of indices as
\[
I=\left\{  (m_1,m_2,m_3,\dots )\mid m_i\in\Z_{\ge0} \hbox{ such that } m_i=0 \hbox{ for } i\gg 0\right\}.
\]
For $\bm{m}=(m_1,m_2,m_3,\dots )\in I$, denote
\[
{\p_\bt}^{\bm m}=\prod_{k\ge1}\left(\frac{\p}{\p {t_k} }\right)^{m_k}, \quad {\p_{\hbt}}^{\bm m}=\prod_{k\ge1}\left(\frac{\p}{\p \hat{t}_k }\right)^{m_k}.
\]
In order to show the equality \eqref{bl}, we only need to check that
the Baker--Akhiezer functions and the adjoint Baker--Akhiezer functions satisfy
\begin{equation}\label{bl-deriv}
\res_{\!z}\left({\p_\bt}^{\bm m}{\p_{\hbt}}^{\bm n}\psi(\bt, \hbt; z)\cdot \psi^\dag(\bt, \hbt; z)\right)=\res_{\!z}\left({\p_\bt}^{\bm m}{\p_{\hbt}}^{\bm n}\hat{\psi}(\bt, \hbt; z)\cdot \hat{\psi}^\dag(\bt, \hbt; z)\right)
\end{equation}
for any indices ${\bm m}, {\bm n}\in I$. In fact, given any ${\bm m}, {\bm n}\in I$, according to Lemma~\ref{thm-bakert} and Proposition~\ref{thm-PPht} there is a pseudo-differential operator $A^{\bm{m}, \bm{n}}\in \cD^-\cap\cD^+$ such that the following two equalities hold simultaneously:
\[
{\p_\bt}^{\bm m}{\p_{\hbt}}^{\bm n}\psi(z)=A^{\bm{m}, \bm{n}}\psi(z), \quad {\p_\bt}^{\bm m}{\p_{\hbt}}^{\bm n}\hat\psi(z)=A^{\bm{m}, \bm{n}}\hat\psi(z).
\]
Then, by using \eqref{baker} and Lemma~\ref{thm-res}, the equality \eqref{bl-deriv} is recast to
\[
\res\left( A^{\bm{m}, \bm{n}}\Phi \Phi^{-1}\right)=\res\left( A^{\bm{m}, \bm{n}}\hat\Phi {\hat\Phi}^{-1}\right),
\]
which is clearly valid. Hence the equality \eqref{bl-deriv} holds
true, and the first assertion is verified.

For the second assertion, one sees that there are uniquely
pseudo-differential operators $\Phi$, $\hat{\Phi}$, $\Psi$ and
$\hat{\Psi}$ such that the functions \eqref{wavef2}--\eqref{wavef2a}
are represented as
\begin{align*}
\psi(\bm{t}, \hat{\bm{t}}; z)=\Phi e^{\xi(\bm{t};z)},& \quad
\hat{\psi}(\bm{t}, \hat{\bm{t}}; z)=\hat{\Phi}  e^{x
z-\xi(\hat{\bm{t}};z^{-1})},
\\
\psi^\dag(\bm{t}, \hat{\bm{t}}; z)=\Psi^*  e^{-\xi(\bm{t};z)},& \quad
\hat{\psi}^\dag(\bm{t}, \hat{\bm{t}}; z)=\hat{\Psi}^*  e^{-x
z+\xi(\hat{\bm{t}};z^{-1})}.
\end{align*}
Moreover, the operators $\Phi$ and $\hat\Phi$ take the form
\eqref{Phi} and \eqref{Phih} respectively, while $\Psi$ and
$\hat\Psi$ take the form \eqref{Phiinv} and \eqref{Phihinv}
respectively. The bilinear equation \eqref{bl} leads to the
following facts.
\begin{itemize}
\item[(i)] For $i\in\Z$, we have (note that when $i<0$ the integral constants on both sides are fixed in the same way)
\[
\res_{\!z}\left(\p^i\psi(\bt, \hbt; z)\psi^\dag(\bt', \hbt'; z)\right)=\res_{\!z}\left(\p^i\hat{\psi}(\bt, \hbt; z)\hat{\psi}^\dag(\bt', \hbt'; z)\right).
\]
Let $(\bt', \hbt')=(\bt, \hbt)$, and then with the help of \eqref{res} we obtain
\begin{align*}
\res \p^i\Phi\Psi=\res \p^i\hat{\Phi}\hat{\Psi}=0,& \quad i{\ge0}; \\
\res \p^i\hat{\Phi}\hat{\Psi}=\res \p^i\Phi\Psi=\dt_{i,-1},& \quad i<0.
\end{align*}
which implies $\Phi\Psi=\hat{\Phi}\hat{\Psi}=1$. So we derive
\begin{equation}\label{}
\Psi= \Phi^{-1} , \quad \hat{\Psi}= \hat{\Phi}^{-1}.
\end{equation}
\item[(ii)]
Denote
\[
X_k=\frac{\p\Phi}{\p t_k}\Phi^{-1}, \quad
\hat{X}_k=\frac{\p\hat{\Phi}}{\p t_k}\hat{\Phi}^{-1}.
\]
Clearly, one has $(X_k)_+=(\hat{X}_k)_-=0$, and that
\begin{align*}
&\frac{\p\psi(z)}{\p t_k}=\left(X_k\Phi +\Phi \p^k\right)e^{\xi(\bm{t};z)}=\left(X_k +\Phi \p^k\Phi^{-1}\right)\psi(z), \\
&\frac{\p\hat{\psi}(z)}{\p t_k}= \left(\hat{X}_k\hat{\Phi}+\dt_{k
1}\hat{\Phi}\p \right) e^{x z-\xi(\hat{\bm{t}};z^{-1})}=
\left(\hat{X}_k+\dt_{k 1}\hat{\Phi}\p\hat{\Phi}^{-1}
\right)\hat{\psi}(z).
\end{align*}
For any $i\in\Z$, we let $\p^i$ act on the derivative of \eqref{bl} with respect to $t_k$, and let $(\bt', \hbt')=(\bt, \hbt)$, then by using \eqref{res} again we obtain
\begin{align*}
\res \p^i\left(X_k +\Phi \p^k\Phi^{-1}\right)\Phi\Phi^{-1}=\res \p^i
\left(\hat{X}_k+\dt_{k 1}\hat{\Phi}\p\hat{\Phi}^{-1}
\right)\hat{\Phi}\hat{\Phi}^{-1}.
\end{align*}
Hence $X_k +\Phi \p^k\Phi^{-1}=\hat{X}_k+\dt_{k
1}\hat{\Phi}\p\hat{\Phi}^{-1} $, and we arrive at
\begin{equation}\label{}
X_k=-\left( \Phi \p^k\Phi^{-1}\right)_-, \quad \hat{X}_k=\left( \Phi
\p^k\Phi^{-1}\right)_+-\dt_{k 1}\hat{\Phi}\p\hat{\Phi}^{-1} .
\end{equation}
Similarly, we can derive
\begin{equation}\label{}
\frac{\p\Phi}{\p\hat{t}_k}\Phi^{-1}=-\left( \hat\Phi \p^{-k}\hat\Phi^{-1}\right)_-, \quad \frac{\p\hat{\Phi}}{\p \hat{t}_k}\hat{\Phi}^{-1}=\left( \hat\Phi \p^{-k}\hat\Phi^{-1}\right)_+.
\end{equation}
\end{itemize}
Taking (i) and (ii) together we achieve the second assertion. The
theorem is proved.
\end{prf}

\section{Additional symmetries versus Virasoro symmetries}

In this section, we want to construct a class of additional
symmetries for the extended KP hierarchy,
following the approach of \cite{OS, Wu-virD}, and then study the
Virasoro symmetries for the constrained KP hierarchies.

\subsection{Additional symmetries for the extended KP hierarchy}
Suppose that the operators $\Phi$ and $\hat{\Phi}$ solve the hierarchy
\eqref{ppt1}, \eqref{ppt2}. Let us introduce two Orlov--Schulman
operators as follows:
\begin{equation}\label{MMh}
M=\Phi\Xi\Phi^{-1}, \quad
\hat{M}=\hat{\Phi}\hat{\Xi}\hat{\Phi}^{-1},
\end{equation}
where
\[
\Xi=\sum_{k\in\Z_{>0}}k t_k  \p^{k-1}, \quad
\hat{\Xi}=x+\sum_{k\in\Z_{>0}} k\hat{t}_k \p^{-k-1}.
\]
Here we assume that all $t_k$ and $\hat{t}_k$ vanish except finitely
many of them, such that the  operators $M$ and $\hat{M}$ are well defined.

\begin{rmk}
There is another way to ensure the Orlov--Schulman operators $M$ and
$\hat{M}$ (even with infinitely many time variables) to be well
defined. Indeed, as what was done in \cite{Wu-virD}, one can extend
the graded algebra $\cA$ to include also  $\{t_k, \hat{t}_k\mid
k\in\Z_{>0}\}$ with $\deg t_k=\deg\hat{t}_k=k$, so that a new graded
algebra  $\tilde{\cA}$ is obtained. Then, the set $\cD^-$ of
pseudo-differential operators of the first type can be extended to
\[
\tilde{\cD}^-=\left\{ \sum_{i\in\Z}\sum_{j\ge \max\{0,i-k\}}a_{i,j} \p^i
\mid a_{i,j}\in\tilde{\cA}_j, k\in\Z \right\}.
\]
So, the operators $M$ and $\hat{M}$ are elements of $\tilde{\cD}^-$
and $\cD^+$ (with $\cA$ replaced by $\tilde{\cA}$) respectively.
\end{rmk}

\begin{lem}\label{thm-Mw}
The Orlov--Schulman operators $M$ and $\hat{M}$ satisfy:
\begin{align}\label{LMLMh}
[P, M]=1,& \quad [\hat{P}^{-1},\hat{M}]=1, \\
M \psi(z)=\frac{\p\psi(z)}{\p z},& \quad \hat{M} \hat{\psi}(z)=\frac{\p\hat{\psi}(z)}{\p z}, \label{Mpsi}  \\
\frac{\pd \dot{M}}{\pd t_k}=[(P^k)_+,\dot{M}], & \quad
 \frac{\pd \dot{M}}{\pd \hat{t}_k}=[-(\hat{P}^k)_-, \dot{M}], \label{Mt}
\end{align}
where  $\dot{M}=M$ or $\hat{M}$.
\end{lem}
\begin{prf}
Based on the definition of $M$ and $\hat{M}$ in \eqref{MMh}, the
first line \eqref{LMLMh} follows from \eqref{PPh}, the second line
\eqref{Mpsi} follows from \eqref{wavef}, while the third line
\eqref{Mt} follows from \eqref{ppt1} and \eqref{ppt2}. The lemma is
proved.
\end{prf}

For any pair of integers $(m,p)$ with $m\ge0$, let
\begin{align}\label{Aml}
& B_{m p}=M^m P^p, \quad
\hat{B}_{m p}=\hat{M}^m \hat{P}^{-p},
\end{align}
and we introduce the following evolutionary equations:
\begin{align}
&\frac{\pd \Phi}{\pd \beta_{m p}}=- (B_{m p})_-\Phi,
\quad \frac{\pd \hat{\Phi}}{\pd \beta_{m p}}=(B_{m p})_+\hat{\Phi}, \label{sml}\\
&\frac{\pd \Phi}{\pd \hat{\beta}_{m p}}=- (\hat{B}_{m p})_-\Phi,
\quad \frac{\pd \hat{\Phi}}{\pd \hat{\beta}_{m p}}=(\hat{B}_{m
p})_+\hat{\Phi}. \label{shml}
\end{align}
As before, such flows are assumed to commute with $\pd/\pd x$.

\begin{lem}\label{thm-MLs}
For any $m, m'\in\Z_{\ge0}$ and $p, p'\in\Z$, the following equalities hold:
\begin{align}
&\frac{\pd \psi(z)}{\pd \dot{\beta}_{m p}}=-(\dot{B}_{m
p})_-\psi(z), \quad \frac{\pd \hat{\psi}(z)}{\pd \dot{\beta}_{m
p}}=(\dot{B}_{m p})_+\hat{\psi}(z), \label{wsdot}
\\
&\frac{\pd P}{\pd \dot{\beta}_{m p}}=[-(\dot{B}_{m p})_-, P], \quad
\frac{\pd \hat{P}}{\pd \dot{\beta}_{m p}}=[(\dot{B}_{m p})_+,
\hat{P}], \label{Lsdot}
\\
&\frac{\pd M}{\pd \dot{\beta}_{m p}}=[-(\dot{B}_{m p})_-, M], \quad
\frac{\pd \hat{M}}{\pd \dot{\beta}_{m p}}=[(\dot{B}_{m p})_+,
\hat{M}], \label{Msdot}
\\
&\frac{\pd B_{m'p'}}{\pd \dot{\beta}_{m p}}=[-(\dot{B}_{m p})_-,
B_{m'p'}], \quad \frac{\pd \hat{B}_{m'p'}}{\pd \dot{\beta}_{m
p}}=[(\dot{B}_{m p})_+, \hat{B}_{m'p'}], \label{Asdot}
\end{align}
where $\dot{\beta}_{m p}=\beta_{m p}, \hat{\beta}_{m p}$ correspond
to $\dot{B}_{m p}=B_{m p}, \hat{B}_{m p}$ respectively.
\end{lem}
\begin{prf}
The equalities \eqref{wsdot}--\eqref{Msdot} follow from the
definition \eqref{sml}, \eqref{shml}. Subsequently,  the equalities \eqref{Asdot} follow from
\eqref{Lsdot} and \eqref{Msdot}. The lemma is
proved.
\end{prf}

\begin{thm}\label{thm-st}
The flows defined by \eqref{sml}, \eqref{shml} commute with those in
\eqref{ppt1}, \eqref{ppt2} that compose the extended KP
hierarchy. More exactly, for any $\dot{\beta}_{m p}=\beta_{m p},
\hat{\beta}_{m p}$ and $\bar{t}_k=t_k, \hat{t}_k$ it holds that
\begin{equation}\label{st}
\left[\frac{\pd}{\pd \dot{\beta}_{m p}}, \frac{\pd}{\pd
\bar{t}_k}\right]\tilde\Phi=0, \quad m\in\Z_{\ge0}, ~~ p\in\Z, ~~ k\in\Z_{>0},
\end{equation}
where $\tilde\Phi=\Phi$ or $\hat\Phi$.
\end{thm}
\begin{prf}
Firstly, from \eqref{PPht}, \eqref{PPhth} and \eqref{Mt} it follows that
\[
\frac{\pd \dot{B}_{m p}}{\pd t_k}=[(P^k)_+,\dot{B}_{m p}],  \quad
 \frac{\pd \dot{B}_{m p}}{\pd \hat{t}_k}=[-(\hat{P}^k)_-, \dot{B}_{m p}],
\]
with $ \dot{B}_{m p}= {B}_{m p}$, $ \hat{B}_{m p}$. Then the
proposition is checked case by case with the help of
Lemmas~\ref{thm-Mw} and \ref{thm-MLs}. For instance,
\begin{align}\label{}
\left[\frac{\pd}{\pd \hat{\beta}_{m p}}, \frac{\pd}{\pd
t_k}\right]\hat{\Phi}=&\frac{\pd}{\pd \hat{\beta}_{m p}} \left(((P^k)_+-\dt_{k1}\hat{P}^{-1} )\hat{\Phi} \right)-\frac{\pd}{\pd
t_k} \left( (\hat{B}_{m p})_+\hat{\Phi} \right) \nn\\
=& [ (P^k)_+-\dt_{k1}\hat{P}^{-1},(\hat{B}_{m p})_+]\hat{\Phi}
 \nn\\
 &+\left([-(\hat{B}_{m p})_-, P^k]_+ - \dt_{k1}[(\hat{B}_{m p})_+,
\hat{P}^{-1}]\right)\hat{\Phi}   -[(P^k)_+,\hat{B}_{m p}]_+\hat{\Phi} \nn \\
=&\left([ (P^k)_+,(\hat{B}_{m p})_+]+[ (P^k)_+,(\hat{B}_{m p})_-]_+
-[ (P^k)_+,\hat{B}_{m p}]_+ \right)\hat{\Phi} =0. \nn
\end{align}
The other cases are similar. Thus the proposition is proved.
\end{prf}

Proposition~\ref{thm-st} means that the flows
\eqref{sml}, \eqref{shml} give a class of symmetries for the extended KP
hierarchy, which are called the \emph{additional
symmetries}.

%\begin{rmk}
%Although $B_{0,2i+1}=2 P^{2i+1}$ (resp. $\hat{B}_{0,-2i-1}=2
%\hat{P}^{2i+1}$), the vector fields $\pd/\pd \beta_{0,2 i+1}$ (resp.
%$\pd/\pd\hat{\beta}_{0,-2 i-1}$) cannot be identified to $2\pd/\pd
%t_{2 i+1}$ (resp. $2\pd/\pd\hat{t}_{2 i+1}$). In fact, they act
%differently on either $M$ or $\hat{M}$.
%\end{rmk}

Let us study the commutation relation between the additional
symmetries themselves. Observe that each commutator $[B_{m p}, B_{m'
p'}]$ is a polynomial in $M$ and $P^{\pm1}$, hence, by virtue of
\eqref{LMLMh}, there exist certain constants $c_{m p, m' p'}^{n q}$
such that
\begin{equation}\label{BB}
[B_{m p}, B_{m' p'}]=\sum_{n, q}c_{m p, m' p'}^{n q}B_{n q}.
\end{equation}
In fact, one has $c_{m p, m' p'}^{n q}=0$ whenever $n\ge m+m'$ or
$|q-(p+p')|>\max(m, m')$, which implies that all but finitely many
structure constants on the right hand side of \eqref{BB} vanish. For
instance, when $m+m'\le2$ one has
\begin{align*}
c_{0p,0p'}^{n q}=0, \quad &c_{0p, 1p'}^{n
q}=p\dt_{n0}\dt_{q,p+p'-1}, \quad
c_{1 p, 1 p'}^{n q}=(p-p')\dt_{n1}\dt_{q, p+p'-1}, \\
&c_{0 p, 2 p'}^{n q}=p(p-1)\dt_{n0}\dt_{q,p+p'-2}+2p \dt_{n1}\dt_{q,
p+p'-1}.
\end{align*}
By virtue of \eqref{LMLMh}, it holds for the same structure
constants that
\begin{equation}\label{BhBh}
[\hat{B}_{m p}, \hat{B}_{m' p'}]=\sum_{n,q}c_{m p, m' p'}^{n
q}\hat{B}_{n q}.
\end{equation}

\begin{prp}\label{thm-ssh}
For the extended KP hierarchy \eqref{ppt1}, \eqref{ppt2}, its additional symmetries defined by \eqref{sml}, \eqref{shml} satisfy:
\begin{align}
&\left[\frac{\pd}{\pd \beta_{m p}}, \frac{\pd}{\pd \beta_{m'
p'}}\right]\tilde\Phi=-\sum_{n,q}c_{m p, m' p'}^{n q}\frac{\pd\tilde\Phi}{\pd \beta_{n
q}},
\\
& \left[\frac{\pd}{\pd \hat{\beta}_{m p}}, \frac{\pd}{\pd
\hat{\beta}_{m'
p'}}\right]\tilde\Phi=\sum_{n,q}c_{m p, m' p'}^{n q}\frac{\pd\tilde\Phi}{\pd \hat{\beta}_{n q}},\\
 &\left[\frac{\pd}{\pd \beta_{m p}}, \frac{\pd}{\pd \hat{\beta}_{m'
p'}}\right]\tilde\Phi=0,
\end{align}
where $\tilde\Phi=\Phi$ or $\hat\Phi$.
\end{prp}
\begin{prf}
The conclusion can be checked case by case with the help of
Lemma~\ref{thm-MLs}. For instance,
\begin{align}\label{}
&\left[\frac{\pd}{\pd \beta_{m p}}, \frac{\pd}{\pd \beta_{m' p'}}\right]\hat{\Phi} \nn\\
=& [ (B_{m'p'})_+, (B_{m p})_+]\hat{\Phi} + [-(B_{m p})_-,
B_{m'p'}]_+\hat{\Phi} - [-(B_{m'p'})_-, B_{m p}]_+\hat{\Phi} \nn \\
=& -[(B_{m p})_+, (B_{m'p'})_+]\hat{\Phi} - [(B_{m p})_-,
(B_{m'p'})_+]_+\hat{\Phi} - [B_{m p}, (B_{m'p'})_-]_+\hat{\Phi} \nn
\\
=&-[B_{m p}, B_{m'p'}]_+\hat{\Phi}
=-\sum_{n,q}c_{m p, m' p'}^{n q}(B_{n q})_+\hat{\Phi} \nn\\
=&-\sum_{n,q}c_{m p, m' p'}^{n q}\frac{\pd\hat{\Phi}}{\pd \beta_{n
q}},
\\
&\left[\frac{\pd}{\pd \hat{\beta}_{m p}}, \frac{\pd}{\pd \hat{\beta}_{m' p'}}\right]\hat{\Phi} \nn\\
=& [ (\hat{B}_{m'p'})_+, (\hat{B}_{m p})_+]\hat{\Phi} + [(\hat{B}_{m
p})_+,
\hat{B}_{m'p'}]_+\hat{\Phi} - [(\hat{B}_{m'p'})_+, \hat{B}_{m p}]_+\hat{\Phi} \nn \\
=&  [(\hat{B}_{m p})_+, (\hat{B}_{m'p'})_-]_+\hat{\Phi} +
[\hat{B}_{m p}, (\hat{B}_{m'p'})_+]_+\hat{\Phi} \nn
\\
=&[\hat{B}_{m p}, \hat{B}_{m'p'}]_+\hat{\Phi} = \sum_{n,q}c_{m p, m'
p'}^{n q}(\hat{B}_{n q})_+\hat{\Phi} \nn
\\
=&\sum_{n,q}c_{m p, m' p'}^{n q}\frac{\pd\hat{\Phi}}{\pd
\hat{\beta}_{n
q}}, \\
&\left[\frac{\pd}{\pd \beta_{m p}}, \frac{\pd}{\pd\hat{\beta}_{m' p'}}\right]\hat{\Phi} \nn\\
=& [ (\hat{B}_{m'p'})_+, (B_{m p})_+]\hat{\Phi} + [(B_{m p})_+,
\hat{B}_{m'p'}]_+\hat{\Phi} - [-(\hat{B}_{m'p'})_-, B_{m p}]_+\hat{\Phi} \nn \\
=& \left([(B_{m p})_+, (\hat{B}_{m'p'})_-]_+ + [(\hat{B}_{m'p'})_-,
(B_{m p})_+]_+\right)\hat{\Phi} =0.
\end{align}
The other cases are checked in the same way. Thus the proposition is
proved.
\end{prf}

This proposition means that the additional symmetries \eqref{sml},
\eqref{shml} for the extended KP hierarchy generate a
$W_{1+\infty}\times W_{1+\infty}$ algebra.

\subsection{Virasoro symmetries for the constrained KP hierarchies}

%In this section, we consider some application of the additional symmetries for the extended KP hierarchy.

Given an arbitrary positive integer $n$, let us consider  the extended KP hierarchy \eqref{ppt1}, \eqref{ppt2} imposed with the following
constraint
\begin{equation}\label{PnPh}
 \Phi\p^n\Phi^{-1}=\hat{\Phi}\p^{-1}\Phi^{-1}.
\end{equation}
Under this constraint, one has $P^n=\hat{P}$ and hence $\p/\p t_{n k}=\p/\p \hat{t}_k$ for
$k\ge1$. Consequently, the extended KP hierarchy is reduced to
% so without loss of generality we will not consider the flows $\p/\p \hat{t}_k$ in this section.
\begin{equation}\label{Lt}
\frac{\p L}{\p t_k}=[(P^k)_+,L], \quad k=1,2,3,\dots,
\end{equation}
where $L:=P^n=\hat{P}$ takes the form
\begin{equation}\label{L}
L=\p^n+u_{1}\p^{n-2}+u_2\p^{n-3}+\dots+u_{n-2}\p+u_{n-1}+(\p-f')^{-1}\rho.
\end{equation}
The system of equations \eqref{Lt} is called the constrained KP hierarchy,
denoted by $\hbox{cKP}_{n,1}$ (see \cite{ANP, BX, Ch, Di}). In fact, if
we write $v=e^f$ and $w=\rho e^{-f}$, then
\begin{equation}\label{}
L_-=v\p^{-1}w
\end{equation}
and it is yielded by the equations \eqref{Lt} that
\begin{equation}\label{}
\frac{\p v}{\p t_k}=\left(P^k\right)_+(v), \quad \frac{\p w}{\p t_k}=-\left(P^k\right)^*_+(w).
\end{equation}

\begin{exa}
The hierarchy
\eqref{Lt} is known as the nonlinear Shr\"{o}dinger hierarchy (see, e.g., \cite{Ch}) when $n=1$, and it is the Yajima--Oikawa hierarchy
\cite{YO} when
$n=2$. For instance, when $n=2$ the operator $L$ takes the form
\[
L=\p^2+u+(\p-f')^{-1}\rho=\p^2+u+v\p^{-1}w,
\]
then the first nontrivial equations
defined by \eqref{Lt} read (cf. \eqref{ekpt2})
\begin{equation}\label{rhophit3}
\frac{\p u}{\p t_2}=2 \rho', \quad \frac{\p\rho}{\p
t_2}=(2\rho f'-\rho')', \quad \frac{\p f}{\p t_2}=u+(f')^2+f'',
\end{equation}
or equivalently,
\begin{equation}\label{rhophit3}
\frac{\p u}{\p t_2}=2 (v w)', \quad \frac{\p v}{\p
t_2}=v''+u v, \quad \frac{\p w}{\p t_2}=-w''-u w,
\end{equation}
which is called the Yajima--Oikawa system.
\end{exa}

We proceed to construct a series of Virasoro symmetries for the
$\hbox{cKP}_{n,1}$ hierarchy \eqref{Lt} with the help of the extended KP hierarchy \eqref{ppt1}, \eqref{ppt2}.
To this end, let us introduce a class of operators as
\begin{equation}\label{Sp}
S_p=\frac{1}{n}M P^{n p+1}+\hat{M}\hat{P}^{p-1},%\in\cD^-\cup\cD^+,
\quad p\in\Z,
\end{equation}
where
\[
M=\Phi\left(\sum_{k\in\Z_{>0}}k t_k\p^{k-1}\right)\Phi^{-1}, \quad  \hat{M}=\hat{\Phi}x\hat{\Phi}^{-1}.
\]
Here the operator $M$ takes the same form as in \eqref{MMh}, while for convenience $\hat{M}$ does not contain the time variables $\hat{t}_k$ as before. One observes that the operators $S_p$ belong to the space $\cD^- +\cD^+$, hence $(S_p)_-\in\cD^-$ and $(S_p)_+\in\cD^+$. The following evolutionary equations are well defined:
\begin{equation}\label{Phis}
\frac{\pd \Phi}{\pd s_p}=-(S_p)_-\Phi, \quad
\frac{\pd\hat{\Phi}}{\pd s_p}=(S_p)_+\hat{\Phi}, \qquad p\ge-1.
\end{equation}

\begin{prp}\label{thm-cKPvir}
The flows defined by \eqref{Phis} are consistent with the constraint \eqref{PnPh}, and they are reduced to
\begin{equation}\label{Ls}
\frac{\p L}{\p s_p}=\left[-(S_p)_-, L\right]=\left[(S_p)_+, L\right], \quad p\ge-1,
\end{equation}
Moreover, the reduced flows defined by  \eqref{Phis} and by \eqref{ppt1}  under the constraint \eqref{PnPh} satisfy:
\begin{align}\label{}
\mathrm{(i)} \qquad & \left[\frac{\p}{\p t_k}, \frac{\p}{\p s_p}\right]L=0,  \\
\mathrm{(ii)} \qquad
& \left[\frac{\p}{\p s_p}, \frac{\p}{\p s_q}\right]L=(q-p)\frac{\p L}{\p s_{p+q}}, \label{ss}
\end{align}
where $k\in\Z_{>0}$ and  $p, q\in\Z_{\ge-1}$.
\end{prp}

\begin{prf}
Since
\begin{equation}\label{LPhi}
L=\Phi\p^n\Phi^{-1}=\hat{\Phi}\p^{-1}\Phi^{-1},
\end{equation}
then the equations in \eqref{Phis} lead respectively to
\begin{equation}\label{Ls2}
\frac{\p L}{\p s_p}=\left[-(S_p)_-, L\right], \quad \frac{\p L}{\p s_p}=\left[(S_p)_+, L\right].
\end{equation}
When $p\ge-1$, it is straightforward to verify
\begin{equation}\label{SL}
[S_p, L]=\frac{1}{n}\left[M P^{n p+1}, P^n\right]+\left[\hat{M}\hat{P}^{p-1}, \hat{P}\right] =-P^{n(p+1)}+\hat{P}^{p+1}=-L^{p+1}+L^{p+1}=0,
\end{equation}
which implies that the equations in \eqref{Ls2} coincide with each other.
Thus the first assertion is obtained.

Let us show the second assertion. On the one hand, by using \eqref{ppt1} we have, for $p\ge-1$ and $k\ge1$,
\begin{align}
\frac{\p S_p}{\p t_k}=& \frac{1}{n}\left[ -(P^k)_-, M P^{n p+1} \right]+\frac{k}{n}P^{k+n p} \nn\\
& +  \left[ (P^k)_+-\dt_{k 1}\hat{P}^{-1}, \hat{M} \hat{P}^{p-1} \right] +\dt_{k 1}\hat{P}^{p-1}\nn\\
=&\frac{1}{n}\left[ -(P^k)_-+P^k, M P^{n p+1} \right]+  \left[ (P^k)_+, \hat{M} \hat{P}^{p-1} \right] \nn\\
=&\left[(P^k)_+, S_p\right],
\end{align}
On the other hand, by using \eqref{Phis} we have, for $p,q\ge-1$,
\begin{equation}\label{}
\frac{\p (M P^p)}{\p s_q}=\left[-(S_q)_-,M P^p\right], \quad \frac{\p (\hat{M} \hat{P}^p)}{\p s_q}=\left[(S_q)_+,\hat{M} \hat{P}^p\right].
\end{equation}
Then the first item is checked as
\begin{align*}
\left[\frac{\p}{\p t_k}, \frac{\p}{\p s_p}\right]L=& \frac{\p}{\p t_k}\left[(S_p)_+, L\right]- \frac{\p}{\p s_p}\left[(P^k)_+, L\right]  \\
=& \left[\left[(P^k)_+, S_p\right]_+, L\right]+ \left[ (S_p)_+, \left[(P^k)_+,L\right]\right] \\
&\quad - \left[\left[-(S_p)_-, P^k\right]_+, L\right]- \left[(P^k)_+, \left[ (S_p)_+,L\right]\right] \\
=& \left[\left[(P^k)_+, (S_p)_+\right], L\right]- \left[ \left[(P^k)_+,L\right],  (S_p)_+\right] - \left[(P^k)_+, \left[ (S_p)_+,L\right]\right]=0.
\end{align*}
For the second item, it is straightforward to calculate
\begin{align*}
\left[\frac{\p}{\p s_p}, \frac{\p}{\p s_q}\right]L=& \frac{\p}{\p s_p}\left[(S_q)_+, L\right]-
\frac{\p}{\p s_q}\left[(S_p)_+, L\right]  \\
=& \left[\left[-(S_p)_-, \frac{1}{n}M P^{n q+1}\right]_+ +\left[(S_p)_+, \hat{M}\hat{P}^{q-1}\right]_+, L\right] \\
&\quad + \left[ (S_q)_+, \left[(S_p)_+,L\right]\right]-(p\leftrightarrow q) \\
=& \Big[\left[-(S_p)_-, \frac{1}{n}M P^{n q+1}\right]_+ +\left[(S_p)_+, \hat{M}\hat{P}^{q-1}\right]_+ -\left[-(S_q)_-, \frac{1}{n}M P^{n p+1}\right]_+  \\
&\quad  -\left[(S_q)_+, \hat{M}\hat{P}^{p-1}\right]_+ + \left[ (S_q)_+,  (S_p)_+\right], L\Big] \\
=&\left[\frac{1}{n^2}X +\frac{1}{n}Y+Z, L\right]
\end{align*}
where, with \eqref{Sp} substituted,
\begin{align*}
X=& \left[-(M P^{n p+1})_-, M P^{n q+1}\right]_+ - \left[-(M P^{n q+1})_-, M P^{n p+1}\right]_+  \\
&\quad  + \left[(M P^{n q+1})_+, (M P^{n p+1})_+\right]  \\
=&  \left[M P^{n q+1} , M P^{n p+1} \right]_+ =n(q-p)(M P^{n (p+q)+1})_+,  \\
Y=& \left[-(\hat{M}\hat{P}^{p-1})_-, M P^{n q+1}\right]_+ + \left[(M P^{n p+1})_+,\hat{M}\hat{P}^{q-1}\right]_+  \\
&- \left[-(\hat{M}\hat{P}^{q-1})_-, M P^{n p+1}\right]_+ -\left[(M P^{n q+1})_+,\hat{M}\hat{P}^{p-1}\right]_+ \\
&+ \left[(M P^{n q+1})_+, (\hat{M}\hat{P}^{p-1})_+\right] + \left[(\hat{M}\hat{P}^{q-1})_+, (M P^{n p+1})_+\right]  \\
=& \left[(M P^{n q+1})_+, \hat{M}\hat{P}^{p-1}, \right]_+ - \left[(M P^{n q+1})_+,\hat{M}\hat{P}^{p-1}\right]_+  \\
 & \quad + \left[(M P^{n p+1})_+,(\hat{M}\hat{P}^{q-1})_+\right]_+    + \left[(\hat{M}\hat{P}^{q-1})_+, (M P^{n p+1})_+\right] =0, \\
Z=&\left[(\hat{M}\hat{P}^{p-1})_+, \hat{M}\hat{P}^{q-1}\right]_+ -\left[(\hat{M}\hat{P}^{q-1})_+, \hat{M}\hat{P}^{p-1}\right]_+ + \left[(\hat{M}\hat{P}^{q-1})_+, (\hat{M}\hat{P}^{p-1})_+\right] \\
=&\left[(\hat{M}\hat{P}^{p-1})_+, \hat{M}\hat{P}^{q-1}\right]_+ -\left[(\hat{M}\hat{P}^{q-1})_+, (\hat{M}\hat{P}^{p-1})_-\right]_+ \\
=&\left[\hat{M}\hat{P}^{p-1} , \hat{M}\hat{P}^{q-1}\right]_+=(q-p)(\hat{M}\hat{P}^{p+q-1})_+.
\end{align*}
So we obtain
\begin{align*}
\left[\frac{\p}{\p s_p}, \frac{\p}{\p s_q}\right]L=(q-p)\left[ (S_{p+q})_+, L\right]=(q-p)\frac{\p L}{\p s_{p+q}}.
\end{align*}
The proposition is proved.
\end{prf}

According to this proposition, the flows \eqref{Ls} give a series of Virasoro symmetries for the  $\hbox{cKP}_{n,1}$ hierarchy \eqref{Lt}. It is worthwhile to indicate that, here the condition $p\ge-1$ is necessary, otherwise \eqref{SL} may not vanish since $L$ has different inverse as an operator in $\cD^-$ or in $\cD^+$.

One observes that the flows $\pd/\pd s_p$ in \eqref{Phis} can be considered as reductions of the linear combinations
\[
\frac{1}{n}\frac{\pd}{\pd
\beta_{1,n p+1}}+\frac{\pd}{\pd \hat{\beta}_{1,-p+1}}
\]
of the additional symmetries \eqref{sml}, \eqref{shml} for the
extended KP hierarchy constrained by \eqref{LPhi}. This observation motivates us to consider whether the symmetries like $\p/\p\beta_{0p}$ or $\p/\p\hat\beta_{0p}$ for the extended KP hierarchy could be reduced to that for the $\hbox{cKP}_{n,1}$ hierarchy.

Similar as in \eqref{Phis}, for arbitrary constants $\ka$ and $\ld$ we can define the following equations:
\begin{equation}\label{Phis2}
\frac{\pd \Phi}{\pd s_p'}=-\left(S_p+(\ka p+\ld)\hat{P}^{p}\right)_-\Phi, \quad
\frac{\pd\hat{\Phi}}{\pd s_p'}=\left(S_p+(\ka p+\ld)\hat{P}^{p}\right)_+\hat{\Phi}, \qquad p\ge-1.
\end{equation}
In the same way as above, by doing the replacement $S_p\mapsto S_p+(\ka p+\ld)\hat{P}^{p}$, we achieve that the flows $\p/\p s_p'$ are consistent with the constraint \eqref{PnPh}, and they are reduced to
\begin{equation}\label{LsL}
\frac{\p L}{\p s_p'}=\left[-(S_p+(\ka p+\ld)\hat{P}^{p})_-, L\right], \quad p\ge-1.
\end{equation}
Moreover, such reduced flows satisfy:
\begin{align}\label{}
\mathrm{(i)} \qquad & \left[\frac{\p}{\p t_k}, \frac{\p}{\p s_p'}\right]L=0,  \\
\mathrm{(ii)} \qquad
& \left[\frac{\p}{\p s_p'}, \frac{\p}{\p s_q'}\right]L=(q-p)\frac{\p L}{\p s_{p+q}'},
\end{align}
where $k\in\Z_{>0}$ and  $p, q\in\Z_{\ge-1}$.

\begin{rmk}
The reduced flows defined by \eqref{Phis} and by \eqref{Phis2} under the constraint \eqref{PnPh} satisfy
\begin{equation}\label{LsLsL}
\left(\frac{\p}{\p s_p'}-\frac{\p}{\p {s}_p} \right) L=\left\{
\begin{array}{cc}
  0,  & p=-1, 0; \\
 -(\ka p+\ld) \left[(L^p)_-, L\right], & p\ge1.
\end{array}\right.
\end{equation}
\end{rmk}

It is known that, for the $\hbox{cKP}_{n,1}$ hierarchy \eqref{Lt}, a series of the Virasoro symmetries was constructed by Aratyn, Nissimov and Pacheva in \cite{ANP} by adding certain  ``ghost'' symmetry flows related to the eigenfunctions characteristic for the operators $L$ and $L^*$ (in fact more general cases were studied there). More exactly, in Proposition~2 of \cite{ANP} the following Virasoro symmetries were given:
\begin{equation}\label{Lstil}
\frac{\p L}{\p \tilde{s}_p}=\left[-\frac{1}{n}(M P^{n p+1})_- +Y_p , L\right], \quad p\ge-1,
\end{equation}
where $Y_{p} =0$ for $p=-1,0,1$, and
\begin{equation}\label{Yp}
Y_{p} =\sum_{j=0}^{p-1} \frac{2 j-p+1}{2}L^{p-j-1}(v)\p^{-1} (L^*)^j(w), \quad p\ge2.
\end{equation}
Here we repeat
\[
v=e^f=\hat{\Phi}(1), \quad w=\rho e^{-f}=\left(\hat{\Phi}^{-1}\right)^*(1),
\]
and note that there are nonlocal-action terms in $Y_p$ whenever $p\ge2$.

\begin{prp}\label{thm-ANP}
The flows defined by \eqref{Lstil} satisfy
\[
\frac{\p L}{\p\tilde{s}_p}=\frac{\p L}{\p s_p'}, \quad  p \ge-1,
\]
where ${\p}/{\p s_p'}$ are given  in \eqref{LsL} with $\ka=-\frac{1}{2}$ and $\ld=\frac{1}{2}$.
\end{prp}
\begin{prf}
Let us fix $\ka=-\frac{1}{2}$ and $\ld=\frac{1}{2}$ in \eqref{LsL}, then
\begin{align*}
\left(\frac{\p}{\p\tilde{s}_p}-\frac{\p}{\p {s}_p'} \right) L
=&\left[-\frac{1}{n}\left(M P^{n p+1}\right)_- +Y_p+\left(S_p-\frac{p-1}{2}\hat{P}^p\right)_-, L\right] = \left[Y_p+Z_p, L\right],
%=&\frac{1}{2}\left[L(v)\p^{-1}w+v\p^{-1}L^*(w), L\right]=\frac{1}{2}\left[ (L^2)_-, L\right].
\end{align*}
where
\begin{equation}\label{Zp}
Z_p:=\left(\hat{M}\hat{P}^{p-1}-\frac{p-1}{2}\hat{P}^p \right)_-=\left(\hat{\Phi}x \p^{-p+1}\hat{\Phi}^{-1}\right)_- - \frac{p-1}{2}\left(\hat{\Phi}\p^{-p}\hat{\Phi}^{-1}\right)_-.
\end{equation}
When $p=-1,0,1$, clearly $Z_p=0=Y_p$, then the flows $\p/\p\tilde{s}_p$ and $\p/\p s_p'$ acting on $L$ coincide. When $p=2$, on one hand from \eqref{Yp} we have
\begin{align*}
Y_2=&\frac{1}{2}\left(-L(v)\p^{-1}w+v\p^{-1}L^*(w)\right) \\
=&-L(v)\p^{-1}w+\frac{1}{2}\left(L(v)\p^{-1}w+v\p^{-1}L^*(w)\right) \\
=&-L(v)\p^{-1}w+\frac{1}{2}\left(L^2\right)_-,
\end{align*}
where the last equality is due to the formula (61) in the appendix of \cite{ANP}, say, in the present case
\[
(L^k)_-=\sum_{j=0}^{k-1}  L^{k-j-1}(v)\p^{-1} (L^*)^j(w), \quad k \ge1.
\]
On the other hand, since
\begin{align*}
\left(\hat{\Phi}x \p^{-1}\hat{\Phi}^{-1}\right)_- =& \hat{\Phi}(x)\cdot \p^{-1} \cdot\left(\hat{\Phi}^{-1}\right)^*(1) =\hat{\Phi}\p^{-1}\hat{\Phi}^{-1}\hat{\Phi}\p(x)\cdot \p^{-1}\cdot \left(\hat{\Phi}^{-1}\right)^*(1)  \\
=&L\hat{\Phi}(1)\cdot\p^{-1}\cdot w=L(v)\p^{-1}w,
\end{align*}
then from \eqref{Zp} we have
\[
Z_2=L(v)\p^{-1}w-\frac{1}{2}\left(L^2\right)_-.
\]
Clearly $Y_2+Z_2=0$, then we obtain $\p L/\p\tilde{s}_2=\p L/\p s_2'$.  When  $p\ge3$, the flows $\p/\p \tilde{s}_p$ and $\p/\p \tilde{s}_p'$ can be determined by those flows with $p=-1,0,1,2$ via the Virasoro commutation relations. Therefore we complete the proof.
\end{prf}

%\begin{rmk}
%The proof of the above proposition implies that the ``ghost'' flows in \cite{ANP} can be illustrated by pseudo-differential operators of the second type. We hope that such illustration would help us to understand the application of the constrained KP hierarchy to the two-matrix model in \cite{ANP-DB}.
%\end{rmk}

\section{Concluding remarks}

In this paper we have represented the extended KP hierarchy into a bilinear equation satisfied by the (adjoint) Baker--Akhiezer functions. What is more, we have constructed a class of additional symmetries for this hierarchy, and studied their reduction properties. Such results, which are analogue to those for the KP and the 2-BKP hierarchies, are expected to extend our understanding to the theory of integrable systems.

At the end of \cite{WZ}, a tau function of the extended KP hierarchy was introduced by using the densities of Hamiltonian functionals. Similar to the KP hierarchy, one can derive a Sato formula that links the tau function to the (adjoint) Baker--Akhiezer functions $\psi(z)$ and $\psi^\dag(z)$. However, the relationship between the tau function and $\hat{\psi}(z)$ (or $\hat{\psi}^\dag(z)$) still needs to be clarified. It is why a bilinear equation of tau function is still missing. We also hope that the tau function of the extended KP hierarchy could be applied in the two-matrix model such as in the context of constrained KP hierarchies \cite{ANP-DB}. This will be studied elsewhere.

\vskip 0.5truecm \noindent{\bf Acknowledgments.} {The authors thank Professor Baofeng Feng for helpful discussions. This work is partially supported by the National
Natural Science Foundation of China Nos.\,12022119, 11771461, 11831017 and 11521101.

\end{document}